\documentclass[reprint,prl,floatfix,twocolumn,superscriptaddress]{revtex4-2}

\usepackage{amsmath}
\usepackage{amssymb}
\usepackage{float}
\usepackage{graphicx}
\usepackage{hyperref}
\usepackage{lipsum}
\usepackage{siunitx}
\usepackage[version=4]{mhchem}

\graphicspath{{../figures/},{../figures_supplement/}}

\makeatletter
\@ifundefined{textcolor}{}
{%
 \definecolor{BLACK}{gray}{0}
 \definecolor{WHITE}{gray}{1}
 \definecolor{RED}{rgb}{1,0,0}
 \definecolor{GREEN}{rgb}{0,1,0}
 \definecolor{BLUE}{rgb}{0,0,1}
 \definecolor{CYAN}{cmyk}{1,0,0,0}
 \definecolor{MAGENTA}{cmyk}{0,1,0,0}
 \definecolor{YELLOW}{cmyk}{0,0,1,0}
}
\usepackage{dcolumn}
\usepackage{bm}
\usepackage{float}
\makeatother

\begin{document}

\title{
Improved Coherence in Optically-Defined Niobium Trilayer Junction Qubits
}

\author{Alexander Anferov}
\email{aanferov@uchicago.edu}
\affiliation{James Franck Institute, University of Chicago, Chicago, Illinois 60637, USA}
\affiliation{Department of Physics, University of Chicago, Chicago, Illinois 60637, USA}

\author{Kan-Heng Lee}
\affiliation{James Franck Institute, University of Chicago, Chicago, Illinois 60637, USA}
\affiliation{Department of Physics, University of Chicago, Chicago, Illinois 60637, USA}

\author{Fang Zhao}
\affiliation{Department of Physics, University of Chicago, Chicago, Illinois 60637, USA}
\affiliation{Fermi National Accelerator Laboratory, 
Batavia, Illinois 60510, USA}

\author{Jonathan Simon}
\affiliation{James Franck Institute, University of Chicago, Chicago, Illinois 60637, USA}
\affiliation{Department of Physics, University of Chicago, Chicago, Illinois 60637, USA}
\affiliation{Department of Applied Physics, Stanford University, Stanford, California 94305, USA}
\affiliation{SLAC National Accelerator Laboratory, Menlo Park, CA, 94025 USA}

\author{David I. Schuster}
\email{dschus@stanford.edu}
\affiliation{James Franck Institute, University of Chicago, Chicago, Illinois 60637, USA}
\affiliation{Department of Physics, University of Chicago, Chicago, Illinois 60637, USA}
\affiliation{Department of Applied Physics, Stanford University, Stanford, California 94305, USA}
\affiliation{SLAC National Accelerator Laboratory, Menlo Park, CA, 94025 USA}

\date{\today }

\begin{abstract}
Niobium offers the benefit of increased operating temperatures and frequencies for Josephson junctions, which are the core component of superconducting devices.
However existing niobium processes are limited by more complicated fabrication methods and higher losses than now-standard aluminum junctions. 
Combining recent trilayer fabrication advancements, methods to remove lossy dielectrics and modern superconducting qubit design, we revisit niobium trilayer junctions and fabricate all-niobium transmons using only optical lithography. 
We characterize devices in the microwave domain, measuring coherence times up to \qty{62}{\us} and an average qubit quality factor above $10^5$: much closer to state-of-the-art aluminum-junction devices. 
We find the higher superconducting gap energy also results in reduced quasiparticle sensitivity above \qty{0.16}{\K}, where aluminum junction performance deteriorates. 
Our low-loss junction process is readily applied to standard optical-based foundry processes, opening new avenues for direct integration and scalability, and paves the way for higher-temperature and higher-frequency quantum devices.
\end{abstract}

\maketitle

\section{Introduction}
A wide variety of superconducting devices have developed on the basis of Josephson junctions: their applications range from quantum-limited amplification and metrology \cite{jeanneret2009metrology,vijay2009jpa,braginski2019electronics} to digital logic \cite{Vernik2017rsfq,Likharev1991rsfq,tolpygo2020rsfq} and they are an attractive platform for scalable quantum computing architectures due to their design flexibility and wide range of coupling strengths.
Increasingly complex and robust quantum circuits have been demonstrated with aluminum junctions \cite{kjaergaard2020qubitReview}, however niobium is a tantalizing alternative superconductor due to its larger energy gap (and thus higher critical temperature and pair-breaking photon frequency) \cite{finnemore1966nbTc}. 
Taking advantage of this wider operating regime, niobium trilayer Josephson junctions became standard for single-flux-quantum circuits operating at liquid helium temperatures \cite{Vernik2017rsfq,Likharev1991rsfq,tolpygo2020rsfq}.
Employing these well-established fabrication processes, some early implementations of superconducting qubits were developed with niobium junctions \cite{dutta2004nb4n,martinis2002nb10n,paik2008nb17ns,weides2011nb400n,kaiser2010nb26n,lisenfeld2007nb2n,yu2004nb24u,yu2005nb10u}.
However, these initial niobium qubits only retained quantum state coherence for less than \qty{400}{\ns}, diminished by coupling to sources of dephasing and dissipation in the junction and the qubit environment.

Minimizing these loss sources is crucial in all sensitive quantum systems, but particularly for qubits, which must remain coherent over the duration of many gate operations.
Significant effort has since been dedicated to investigating and reducing sources of decoherence \cite{mcdermott2009materialDecoherence}, demanding either adjustments of circuit geometry to limit or dilute coupling to spurious channels, or reducing the use of lossy amorphous dielectric materials.
The need for insulated wiring contacts in these niobium trilayer junctions required growing passivating amorphous dielectric material in direct contact with the junction barrier, which likely degraded early qubit coherence \cite{verjauw2021nbOxide,Altoe2022NbOx}, and limited their use in quantum devices.
Higher temperature junctions with low loss promise a transformative source of strong nonlinearity for high-frequency quantum devices \cite{anferov2020mm4WM,kumar2022mmTransduction}, and have since seen renewed interest from efforts to integrate digital and quantum logic \cite{liu2023sfqControl,leonard2019sfqControl,mcdermott2014sfqControl}, and the exploration of tunnel barrier materials beyond the limitations of aluminum \cite{zhao2020nbMerge,nakamura2011nbn400n,yu2002nbn4u}. 
Notably, by removing amorphous insulating scaffolding and increasing the circuit volume to reduce junction participation, qubits with epitaxially grown NbN junctions with crystalline AlN barriers have increased coherence times to \qty{16}{\us} \cite{kim2021nbn16u}.
We apply similar improvements to traditional Nb/Al/AlO$_x$ processes, which are attractive due to the simpler deposition methods required. 

\begin{figure*}
\centering
\includegraphics[width=6.69in]{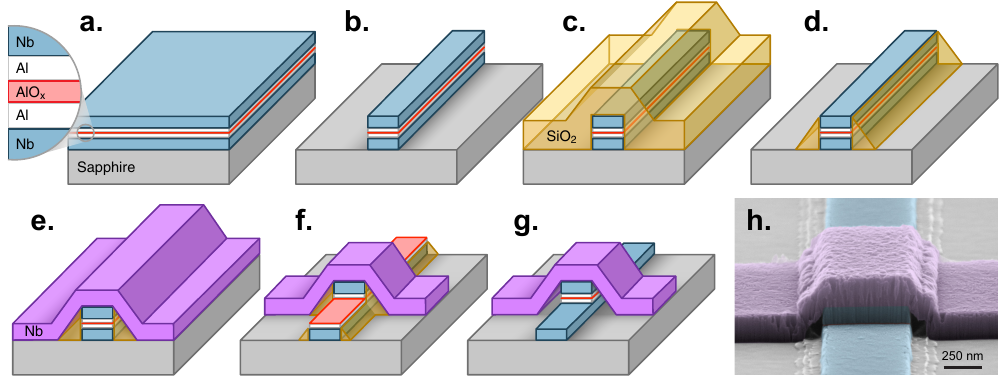}
\caption{
Junction fabrication process. (a) Trilayer is deposited and oxidized in-situ. (b) First layer is etched with a chlorine RIE. (c) \ce{SiO2} is grown isotropically. (d) Sidewall spacer is formed by anisotropic etching with fluorine chemistry. (e) Surface oxides are cleaned in vacuum and wiring layer (purple) is deposited. (f) Second junction finger (and other circuit elements) are defined by a fluorine plasma etch selective against Al. (g) Final devices undergo a wet etch to further remove \ce{SiO2}, exposed Al and some \ce{NbO_x}. (h) Color-enhanced electron micrograph of a finished trilayer junction with dimensions $500\times600~\text{nm}$.
\label{fig:fig1}}
\end{figure*}

In this letter, we use an improved fabrication method to revisit niobium trilayer junctions as the core component of transmon qubits and explore their coherence properties. 
We describe a method to form a temporary self-aligned sidewall-passivating spacer structure based on Ref. \cite{gronberg2017swaps}, which limits the amorphous spacer material to the smallest necessary region, and can later be chemically removed to further reduce dielectric loss. 
We find that high-temperature spacer growth methods greatly reduce the critical current density of the junction barrier, allowing us to utilize exclusively optical lithography to fabricate high-nonlinearity junctions for microwave qubits. 
We find that our all-niobium qubits have lifetimes as high as \qty{62}{\us} with an average qubit quality factor of $2.57\times10^5$: much closer to state-of-the-art qubits than past Nb/Al/AlO$_x$ devices \cite{dutta2004nb4n,martinis2002nb10n,paik2008nb17ns,weides2011nb400n,kaiser2010nb26n,lisenfeld2007nb2n}.
We further observe that the higher superconducting gap energy results in reduced sensitivity to quasiparticles, particularly above \qty{160}{\milli\kelvin}, where conventional aluminum-junction qubit performance deteriorates \cite{paik2011qubit3d,reagor2016qubit,connolly2023nonequilib}. 
These results demonstrate the reemergent relevance of niobium junctions for pushing the boundaries of superconducting devices.

\section{Trilayer Fabrication}
Despite niobium’s attractive electrical properties, in thin layers its oxides are imperfect insulators with high dielectric loss \cite{verjauw2021nbOxide,Altoe2022NbOx}, resulting in very poor natural tunnel junction barriers.
Aluminum, on the other hand, forms a thin self-terminating oxide with low leakage and loss, but has a low critical temperature.
The trilayer method leverages the strengths of both of these materials by using a thin layer of oxidized aluminum as the tunnel barrier and encapsulating it with niobium: through the proximity effect the Josephson junction inherits desired electrical properties and a clean tunnel barrier.
This trilayer structure is typically grown on a wafer-scale as the first step in fabrication, enabling excellent uniformity \cite{Tolpygo2015nbplanarized,bumble2009submicrometer} and high purity growth methods. 

Our fabrication process (see Appendix \ref{appendix:a}) is illustrated in Fig. \ref{fig:fig1}.
Similar to methods using sputtering, our trilayer is formed in a shadow-evaporation-compatible electron-beam system by  depositing \qty{80}{\nm} of Nb and \qty{8}{\nm} of Al on high-purity single-crystal sapphire that has been annealed and chemically etched to remove surface damage. 
The deposition rate is kept high to maximize film quality (see Appendix \ref{appendix:b}). 
To reduce defects and promote aluminum oxide formation \cite{braginski1986crystalox}, the aluminum is first ion milled then oxidized with an \ce{O2}-\ce{Ar} mixture. 
To prevent oxygen diffusion into the Nb layer and the formation of lossy \ce{NbO_x} \cite{tolpygo2010diffstop,verjauw2021nbOxide}, the oxidized Al surface is protected by a thin (\qty{3}{\nm}) capping layer of Al. 
This layer is deposited while rotating the substrate at an angle for complete coverage while keeping it thin enough to avoid affecting junction properties.
A (\qty{150}{\nm}) thick counter electrode is then deposited on top, forming the trilayer in-situ, without breaking vacuum.

The trilayer is patterned with I-line \footnote{exposed with a \qty{375}{\nm} laser} photolithography and plasma-etched in one step with \ce{Cl2}, \ce{BCl3} and \ce{Ar} to define the bottom electrode (Fig. \ref{fig:fig1}b). 
As it is necessary to make contact to the counter electrode without touching the base electrode, we then form an insulating sidewall-passivating spacer structure \cite{gronberg2017swaps}.
Amorphous \ce{SiO2} is grown isotropically (Fig. \ref{fig:fig1}c) by either plasma-enhanced chemical vapor deposition (PECVD) which heats the wafer to \qty{300}{\degreeCelsius} for \qty{16}{\min} or high density plasma-enhanced chemical vapor deposition (HDPCVD) (\qty{90}{\degreeCelsius}).
The \ce{SiO2} is now etched anisotropically with a highly directional \ce{CF4}, \ce{CHF3} and \ce{Ar} plasma, which forms the spacer structure when the bulk material has been etched away (Fig. \ref{fig:fig1}d).
The contaminated trilayer surface is ion milled, and the \qty{160}{\nm} Nb wiring layer is electron-beam-deposited on the sample (Fig. \ref{fig:fig1}e). 
We verify that this forms a low-resistance contact to the counter electrode (see Appendix \ref{appendix:b}).

The wiring layer is patterned and a selective \ce{SF6}, \ce{CHF3}, \ce{O2} and \ce{Ar} plasma etch removes the wiring layer and the counter electrode, defining the perpendicular top junction electrode (Fig. \ref{fig:fig1}f).
This etch is carefully optimized to minimize the formation of lossy fluorocarbon polymers \cite{coburn1979plasmaEtching} (see Appendix \ref{appendix:c}) while preserving chemical selectivity: and although the plasma etches the Al layers far slower than Nb, the etch is still timed to finish a few seconds after the counter electrode is fully removed to limit excessive polymer deposition.
Finally, to further remove the lossy amorphous materials present in the junction, a solution of \ce{NH4F} and acetic acid \footnote{Transene AlPAD Etch 639} are used to dissolve the remaining \ce{SiO2}: this process additionally removes any exposed Al and a small amount of surface \ce{NbO_x} (Fig. \ref{fig:fig1}g).
As this step can dissolve aluminum in the junction as well, etch times are kept below \qty{15}{\s}. 
This final treatment could likely be improved with a \ce{HF} vapor etch, which has shown good results forming similar contact structures \cite{dunsworth2018airbridge}.

\begin{figure}
\centering
\includegraphics[width=3.37in]{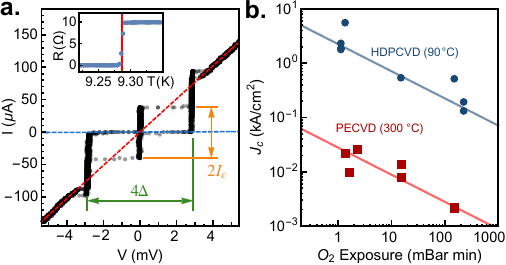}
\caption{
Junction properties. (a) Current-voltage relations for an un-shunted junction at \qty{860}{\milli\kelvin} with $I_c=\qty{38}{\micro\A}$ and an energy gap $2\Delta=\qty{2.89}{\milli\electronvolt}$.
Bulk resistivity measurements (inset) give a critical temperature of $T_c=\qty{9.28}{\K}$.
Above \qty{4}{\milli\V}, a linear fit (red dashed line) gives $R_n=\qty{39}{\ohm}$, and a fit to the sub-gap region (blue dashed line), estimates sub-gap resistance $R_s>\qty{8}{\kohm}$. 
(b) Critical current density $J_c$ (found by fitting room-temperature junction resistance as a function of junction area) as a function of oxygen exposure $E$ measured for various wafers made with two deposition processes. The expected empirical $E^{-1/2}$ relationships are plotted as guides to the eye.
\label{fig:fig2}}
\end{figure}

\section{Junction Properties}
We verify the expected Josephson junction behavior \cite{stewart1968ivcurves} in our devices by measuring their hysteretic current-voltage curves in Fig. \ref{fig:fig2}a. 
When cooled to \qty{860}{\milli\kelvin}, the un-shunted junction shows a zero-resistance superconducting branch up to the critical current $I_c$, and an energy gap $2\Delta=\qty{2.89}{\milli\electronvolt}$. 
By comparing this value with critical temperature measured with resistivity, we find a relationship $2\Delta/k_B T_c=3.61$: slightly lower than reported values for pure Nb \cite{novotny1975nbGap,turneaure1968nbGap}.
Measuring the asymptotic normal state resistance $R_n$ above the energy gap we find a $I_c R_n$ product of \qty{1.5}{\milli\V}, similar to values reported previously for Nb trilayer junctions \cite{morohashi1987nbjrev,gronberg2017swaps,tolpygo2010nbVariations,Tolpygo2015nbplanarized,bumble2009submicrometer}.
Although measurements of the subgap region were limited by the experiment hardware, no excessive subgap leakage currents are observed.

Using the $I_c R_n$ product found above, we can use room-temperature junction resistances to predict low-temperature properties \cite{golubov2004phaseRelations,ambegaokar1963relation}.
Fitting the measured resistance for junctions of varying areas with two free parameters, specific resistivity and junction critical dimension bias (see Appendix \ref{appendix:d}), we obtain the effective junction areas and the specific critical current density $J_c$ for each wafer.
This method allows us to easily investigate effects of the fabrication process on junction electrical parameters.
For Nb trilayers, the critical current density is sensitive to temperature \cite{morohashi1987nbjrev} as well as oxygen exposure $E$, the product of oxygen partial pressure and oxidation time: this relationship has been empirically found to match $J_c\propto E^{-0.5}$ \cite{morohashi1987nbjrev,wen2009o2density,gronberg2017swaps,sugiyama1995o2dependence,kleinsasser1995o2dependence}.
In Fig. \ref{fig:fig2}b we plot $J_c$ as a function of $E$ for wafers with trilayers grown using various oxidation parameters and fabricated with two spacer deposition methods.
For the HDPCVD junctions, we find critical current densities in the \unit{\kilo\A\per\cm\squared} range, comparable with other methods \cite{gronberg2017swaps,anders2009nbCrossJJ,morohashi1987nbjrev,tolpygo2010nbVariations}, and observe reasonable agreement with the oxygen exposure dependence described above.
The effect of process temperature is readily apparent when we examine junctions with high-temperature-grown PECVD spacers: compared to HDPCVD junctions, we observe nearly a factor of 50 reduction in $J_c$.
We find this temperature-annealing effect activates above \qty{200}{\degreeCelsius} (see Appendix \ref{appendix:e}), in agreement with \cite{morohashi1987nbjrev},
and is likely the result of reduced barrier transparency \cite{senapati2010pinhole} from diffusion.

\section{Microwave Qubits}
\begin{figure*}
\centering
\includegraphics[width=6.69in]{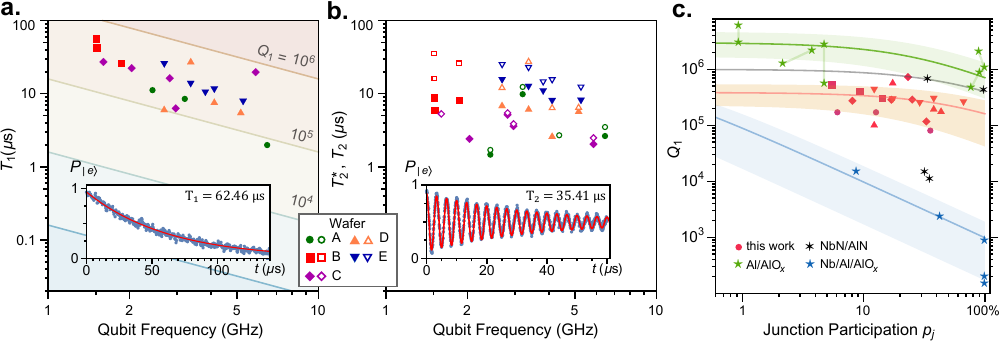}
\caption{
Qubit Properties. 
(a) Average qubit decay time $T_1$ extracted by fitting the exponential decay of excited state population in (b) plotted as a function of qubit frequency, grouped by wafer. Lines indicate qubit quality factor $Q_1=\omega_q T_1$. We find an overall mean $Q_1$ of $2.57\times10^5$ with some wafer to wafer variation. 
(c) Ramsey dephasing time $T_2^*$ (filled points) and Hahn-echo dephasing time $T_2$ (hollow points) extracted by fitting the exponential decay of oscillations in (d) as a function of qubit frequency. 
We find an overall average $T_2^*$ and $T_2$ of $6.643~\mu$s and $12.916~\mu$s respectively.
(e) Qubit quality factors as a function of their junction participation ratio plotted for devices in this work (reds) and in literature (blue, black, green). Lines and shaded confidence regions show $Q_1^{-1}=p_j/Q_j+p_0/Q_0$ as a guide to the eye.
\label{fig:fig3}}
\end{figure*}
With access to wide ranges of $J_c$, we can use PECVD-annealed junctions to realize qubit junctions with areas between $0.16$--$1.1~\mu\text{m}^2$, which are large enough for optical lithography: while lower resolution than electron beam lithography, this speeds up fabrication for devices with large numbers of junctions, and enables seamless integration with superconducting digital logic processes \cite{tolpygo2020rsfq}.
To investigate the junction performance in the context of quantum devices, we fabricate transmon qubits \cite{koch2007cpb} in the well-studied microwave regime (1--8~GHz).
Aside from the junction we use a standard qubit geometry \cite{barends2013xmon} (see Appendix \ref{appendix:f}) capacitively coupled to a coplanar waveguide resonator for dispersive readout. 
The qubit capacitor, ground plane and readout resonator are defined on either the base electrode or wiring layer, so no additional fabrication steps are needed.
Chips with several qubits and their readout resonators sharing a common microwave feedline are characterized at the base stage of a dilution refrigerator (\qtyrange[range-units = single,range-phrase = --]{45}{95}{\milli\K}). 
Using microwave spectroscopy \cite{koch2007cpb} we verify our qubits have anharmonicities around \qty{140}{\MHz}, and couple to their readout resonators with bare coupling strengths $g/h=30$--60~MHz.

\begin{figure}[b]
\centering
\includegraphics[width=3.37in]{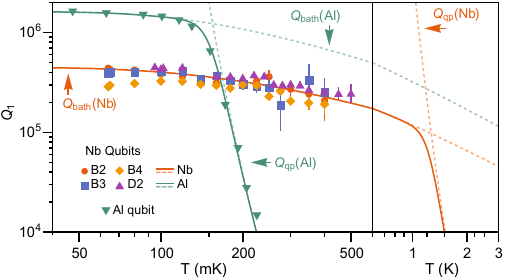}
\caption{
Qubit quality factors from wafers B, D as a function of temperature. 
A mild decrease is observed at higher temperatures consistent with the system bath temperature $Q_\text{bath}$, however lifetimes are virtually unaffected by quasiparticles $Q_\text{qp}$ (red lines). 
We also plot quality factors of an Al junction qubit, whose performance is noticeably limited by tunneling quasiparticles above \qty{160}{\milli\kelvin} (green lines), whereas the Nb junction wouldn't see an effect until about \qty{1.1}{\kelvin}.
\label{fig:fig4}}
\end{figure}

For superconducting qubits, the relaxation time and dephasing time are parameters of particular interest, as they dictate qubit limitations and act as sensitive probes for loss channels.
We measure relaxation time by placing each qubit in its excited state and measuring it after time $t$: fitting the exponential decay gives the characteristic time $T_1$.
We perform these measurements for each qubit and show averaged results as a function of qubit frequency in Fig. \ref{fig:fig3}a, finding $T_1=\qty{62.4}{\us}$ for our best device.
To probe loss channels in detail we use the frequency-independent qubit quality factor $Q_1=\omega_q T_1$, which we find for our devices is on average above $10^5$: within an order of magnitude of recent aluminum qubits \cite{mamin2021alMerge,saxberg2022mott,place2021ta,premkumar2021nb,kjaergaard2020qubitReview} and similar to readout resonator quality factors (see Appendix \ref{appendix:g}).
We also perform a Ramsey experiment to measure the dephasing time $T_2^*$, and a Hahn-echo experiment to characterize the spin-echo dephasing time $T_2$.
We find that $T_2^*$ is within a factor of 2 of $T_1$, and particularly limited for lower-frequency qubits (particularly below \qty{2}{\GHz}) which experience increased environment noise, but also have increased charge sensitivity \cite{koch2007cpb,serniak2019chargeparity} leading to higher noise-dependent variation in qubit frequency.
While this increases dephasing, these devices could be useful for investigating quasiparticle dynamics \cite{serniak2019chargeparity,catelani2014quasiparticle} in Nb.
The $T_2$ values, which decouple slow frequency drifts are noticeably higher, demonstrated in particular by the charge-sensitive qubits from wafer B.
This suggests that along with improved filtering, increasing the $J_c$ for low-frequency qubits could improve dephasing by reducing charge dispersion \cite{koch2007cpb}.

Our qubits have relatively large junctions compared to typical designs \cite{place2021ta,premkumar2021nb,saxberg2022mott,mamin2021alMerge} making them more sensitive to junction coherence properties.
Using the energy participation ratio \cite{mamin2021alMerge,wang2015psurf} of the junction $p_j = C_j/C_\Sigma$ in our devices, we can separate the loss contributions from the junction $Q_j$ independent of other decoherence channels $Q_0$, expressing the qubit quality factor as a weighted sum $Q_1^{-1}=p_j/Q_j + p_0/Q_0$.
We summarize qubit coherence properties with respect to their junction participation in Fig. \ref{fig:fig3}c.
For our devices (red), we estimate an effective junction quality factor of $10^5$: approximately 100 times greater than previous Nb/Al/$\text{AlO}_\text{x}$ qubits (blue) \cite{dutta2004nb4n,martinis2002nb10n,paik2008nb17ns,weides2011nb400n,kaiser2010nb26n,lisenfeld2007nb2n,yu2004nb24u,yu2005nb10u}, and much closer to epitaxial NbN junctions (black) \cite{nakamura2011nbn400n,yu2002nbn4u,kim2021nbn16u} and modern aluminum-junction qubits (green) \cite{place2021ta,premkumar2021nb,saxberg2022mott,mamin2021alMerge}.
Extrapolating to lower $p_j$ values, we find our device loss is largely not limited by the junction, indicating that material refinements and device engineering could further improve qubit performance.

Compared to aluminum, niobium's higher gap energy leads to reduced device sensitivity to thermal quasiparticles \cite{mattis1958bardeen}, and shorter quasiparticle lifetimes \cite{leo2011nbquasiparticletime}, which may help reduce the impact of non-equilibrium quasiparticles on qubit lifetimes \cite{connolly2023nonequilib,martinis2009nonequilib,catelani2011quasiparticle,catelani2014quasiparticle}.
To investigate this thermal resilience, we measure our qubits at increased operating temperatures, shown in Fig \ref{fig:fig4}. 
We observe a mild decrease in $T_1$ with temperature above 160 mK, consistent with heating from the environment bath \cite{lisenfeld2007nb2n}, but importantly don’t see the drastic temperature dependence expected for qubit loss induced by tunneling quasiparticles \cite{catelani2011quasiparticle,catelani2014quasiparticle}, in line with expectations for niobium.
The advantage of higher-temperature junctions is apparent when comparing our qubit performance to an aluminum counterpart: above \qty{160}{\milli\K}, the aluminum qubit is quickly overwhelmed with quasiparticle-induced decoherence, whereas for our devices, $T_1$ is relatively unchanged.
At such elevated temperatures, both devices face challenges from increased thermal microwave noise, motivating alternate qubit architectures \cite{Earnest2018heavyflux,Teoh2023duorail} or higher qubit frequencies to explore this regime.

\section{Conclusion}
We have described a Nb/Al/$\text{AlO}_\text{x}$/Al/Nb trilayer fabrication method demonstrating a 100-fold improvement in junction loss at the single-photon level.
By removing lossy dielectric materials wherever possible, we use our low current density junction process to fabricate microwave transmon qubits using I-line photolithography demonstrating qubit quality factors within an order of magnitude of recent aluminum devices.
Our qubits have relatively high junction participation ratios, which could either be reduced with smaller junctions (defined by electron-beam lithography) to improve coherence, or exploited further to significantly reduce qubit size \cite{zhao2020nbMerge,mamin2021alMerge}.

Together with this device footprint flexibility, our all-optical qubit process opens the door to large-scale direct integration of scalable quantum processors with digital superconducting logic \cite{liu2023sfqControl,leonard2019sfqControl,mcdermott2014sfqControl}.
Niobium's higher energy gap significantly reduces sensitivity to quasiparticles for our junctions compared to aluminum analogues, allowing operation at much higher frequencies, and resulting in longer relaxation times above \qty{160}{\milli\K} where conventional qubit properties deteriorate.
Combined with their low loss, which could be further reduced through material optimization \cite{verjauw2021nbOxide,Altoe2022NbOx,premkumar2021nb}, these properties make our trilayer junctions a promising candidate for quantum architectures with lower cooling power requirements, hybrid qubit systems requiring elevated temperatures, and enable new possibilities for nonlinear elements at millimeter-wave frequencies \cite{anferov2020mm4WM,kumar2022mmTransduction}, paving the way for higher temperature, higher frequency quantum devices.

\section{Acknowledgements}

The authors thank P. Duda and J. Martinez for assistance with fabrication development, and S. Anferov for assistance with air-free techniques.
We acknowledge useful discussions with G. Catelani.
This work is supported by the U.S. Department of Energy Office of Science National Quantum Information Science Research Centers as part of the Q-NEXT center, and partially supported by the University of Chicago Materials Research Science and Engineering Center, which is funded by
the National Science Foundation under Grant No. DMR-1420709.
This work made use of the Pritzker Nanofabrication Facility of the Institute for Molecular Engineering at the University of Chicago, which receives support from Soft and Hybrid Nanotechnology Experimental (SHyNE) Resource (NSF ECCS-2025633).

\appendix

\section{Fabrication Methods}
\label{appendix:a}
\begin{table*}[hbt]
\noindent
\begin{center}
\begin{tabular}{|c||c|c|c|c|c|c|c|c|c|c|c|c|}
\hline
&T(\qty{}{\degreeCelsius})&Pressure &
ICP/Bias Power& \ce{Cl2} & \ce{BCl3} &\ce{Ar} &\ce{CF4} &\ce{CHF3} &\ce{SF6} &\ce{O2} &etch time &etch rate\\
\hline
Etch 1 (Fig. \ref{fig:fig1}b) &$20\pm0.1$ &5 mT &\qty{400}{\watt} / \qty{50}{\watt} &30 &30 &10 &- &- &- &- &50-\qty{60}{\s}&$\sim\qty{4.5}{\nm/\s}$\\
Etch 2 (Fig. \ref{fig:fig1}d) &$20\pm0.1$ &30 mT &\qty{500}{\watt} / \qty{60}{\watt} &- &- &10 & 30& 20&- &- &120-\qty{140}{\s} s&$\sim\qty{2}{\nm/\s}$\\
Etch 3 (Fig. \ref{fig:fig1}f) &$20\pm0.1$ &5 mT &\qty{400}{\watt} / \qty{60}{\watt}  &- &- &7 &- &20 &40 &4 &65-\qty{90}{\s}&$\sim\qty{4.5}{\nm/\s}$\\
\hline
\end{tabular}
\end{center}
\caption{Plasma etch parameters used in the ICP-RIE etches described in the process. Etches are performed in an Apex SLR ICP etcher. Gas flows are listed in sccm.}
\label{tab:etches}
\end{table*}

C-plane polished sapphire wafers are ultrasonically cleaned in toluene, acetone, methanol, isopropanol and de-ionized (DI) water, then etched in a piranha solution kept at \qty{40}{\degreeCelsius} for 2 minutes and rinsed with de-ionized water.
Immediately following, the wafers are loaded into a Plassys MEB550S electron-beam evaporation system, where they are baked by heating the stage to $>$\qty{200}{\degreeCelsius} under vacuum for an hour to help remove water and volatiles.
When a sufficiently low pressure is reached ($<\qty{5e-8}{\milli\Bar}$), titanium is electron-beam evaporated to bring the load lock pressure down even further.
The trilayer is now deposited by first evaporating \qty{80}{\nm} of Nb at $>\qty{0.5}{\nm}\text{/s}$ while rotating the substrate.
After cooling for a few minutes, \qty{8}{\nm} of aluminum is deposited while rotating the substrate at a shallow angle (10 degrees) to improve conformality.
The aluminum is lightly etched with a \qty{400}{\V}, \qty{15}{\mA} \ce{Ar+} beam for \qty{10}{\s}, then oxidized with a mixture of $15\%$ \ce{O2}:\ce{Ar} at a static pressure between 2--50~mBar for 1.5--40~min.
After pumping to below ($<\qty{e-7}{\milli\Bar}$), titanium is again used to bring the vacuum pressure down to the low $\qty{e-8}{\milli\Bar}$ range.
We note that the pressure for the remainder of the trilayer deposition is higher than for the first Nb layer.
The second \qty{3}{\nm} layer of Al is evaporated vertically while rotating the substrate to minimize void formation in the following layer.
The counterelectrode is then formed by evaporating \qty{150}{\nm} of Nb at $>\qty{0.5}{\nm}\text{/s}$.
The substrate is allowed to cool in vacuum for several minutes, and we attempt to form a thin protective coating of pure \ce{Nb2O5} by briefly oxidizing the top surface at $\qty{3}{\milli\Bar}$ for \qty{30}{\s}.

The wafers are mounted on a silicon handle wafer using AZ1518 photoresist cured at \qty{115}{\degreeCelsius}, then coated with \qty{1}{\um} of AZ MiR 703 photoresist and exposed with a \qty{375}{\nm} laser in a Heidelberg MLA150 direct-write system.
The assembly is hardened for etch resistance by a \qty{1}{\min} bake at \qty{115}{\degreeCelsius} then developed with AZ MIF 300, followed by a rinse in DI water.
The entire trilayer structure is now etched in a chlorine inductively coupled plasma reactive ion etcher (Etch 1 in Table \ref{tab:etches}).
The plasma conditions are optimized to be in the ballistic ion regime, which gives high etch rates with minimal re-deposition.
Immediately after exposure to air, the wafer is quenched in DI water: this helps prevent excess lateral aluminum etching by quickly diluting any surface \ce{HCl} (formed by adsorbed \ce{Cl} reacting with water vapor in the air).
The remaining photoresist is thoroughly dissolved in a mixture of \qty{80}{\degreeCelsius} n-methyl-2-pyrrolidone with a small addition of surfactants, which also removes the substrate from the handle wafer.

The wafer is ultrasonically cleaned with acetone and isopropanol, then \ce{SiO2} spacer is grown by either HDPCVD or PECVD.
For PECVD, \ce{SiH4} and \ce{N2O} are reacted in a \qty{100}{\watt} plasma with the chamber at \qty{300}{\degreeCelsius}.
The complete process (including chamber cleaning pumping and purging steps) takes approximately 15 minutes.
For HDPCVD, the wafer is mounted on a silicon handle wafer using Crystalbond 509 adhesive softened at \qty{135}{\degreeCelsius}, then the spacer is deposited with a \ce{SiH4} \ce{O2} and \ce{Ar} plasma, with the substrate heated to \qty{90}{\degreeCelsius}. 
The wafers are now etched in a fluorine reactive ion etch (Etch 2 in Table \ref{tab:etches}).
This etch is optimized to be directional but in the diffusive regime to promote chemical selectivity while enabling the formation of the spacer structure.
At this point minimizing oxide formation is crucial since the top surface of the trilayer is exposed and will need to form a good contact to the wiring layer, so immediately following the completion of the etch, wafers are separated from the handle wafer by heating to \qty{135}{\degreeCelsius}, ultrasonically cleaned of remaining adhesive in \qty{40}{\degreeCelsius} acetone and isopropanol, then immediately placed under vacuum in the deposition chamber, where they are gently heated to \qty{50}{\degreeCelsius} for 30 min to remove remaining volatiles.

The contaminated and oxidized top surface of the counter electrode is etched with a \qty{400}{\V}, \qty{15}{\mA} \ce{Ar+} beam for \qty{5}{\min}, which is sufficient to remove any residual resistance from the contact.
After pumping to below ($<\qty{e-7}{\milli\Bar}$), titanium is used to bring the vacuum pressure down to the low $\qty{e-8}{\milli\Bar}$ range.
The wiring layer is now formed by evaporating \qty{160}{\nm} of Nb at $>\qty{0.5}{\nm}\text{/s}$.
The substrate is allowed to cool in vacuum for several minutes, and the wiring layer is briefly oxidized with $15\%$ \ce{O2}:\ce{Ar} at $\qty{3}{\milli\Bar}$ for \qty{30}{\s} to promote a thin protective coating of pure \ce{Nb2O5}.
The wafers are again mounted on a handle wafer, coated with AZ MiR 703 photoresist and exposed with a \qty{375}{\nm} laser.
The assembly is hardened for etch resistance by a \qty{1}{\min} bake at \qty{115}{\degreeCelsius} before development.
The final structure is now defined with a fluorine reactive ion etch (Etch 3 in Table \ref{tab:etches}).
This step proves to be highly problematic as it easily forms inert residues, and needs to be highly chemically selective in order to avoid etching through the aluminum, so the plasma is operated in a low-density ballistic regime with the addition of \ce{O2} which helps passivate exposed aluminum and increase selectivity.
The etch time is calculated for each wafer based on visual confirmation when the bare wiring layer is etched through.
We remove crosslinked polymers from the photoresist surface with a mild \qty{180}{\watt} room temperature oxygen plasma that minimally oxidizes the exposed Nb (though find this is not very effective).
The remaining resist is now fully dissolved in \qty{80}{\degreeCelsius} n-methyl-2-pyrrolidone with surfactants.

With the junctions now formed, the wafer is ultrasonically cleaned with acetone and isopropanol, coated with a thick protective covering of photoresist (MiR 703) cured at \qty{115}{\degreeCelsius}, and diced into \qty{7}{\mm} chips.
The protective covering is now dissolved in \qty{80}{\degreeCelsius} n-methyl-2-pyrrolidone with surfactants (we find this can also help remove stubborn organic residues from previous steps), and the chips are given a final ultrasonic clean with with acetone and isopropanol.
The remaining silicon spacer is now dissolved by a short 10-15$~$s etch in a mixture of ammonium fluoride and acetic acid (AlPAD Etch 639), quenched in de-ionized water, then carefully dried from isopropanol to preserve the now partially suspended wiring layer.
The finished chips are packaged and cooled down within a couple hours from this final etch to minimize any \ce{NbO_x} regrowth from air exposure.

\section{Junction Superconductor Properties}
\label{appendix:b}
\begin{figure*}[ht]
\centering
\includegraphics[width=6.69in]{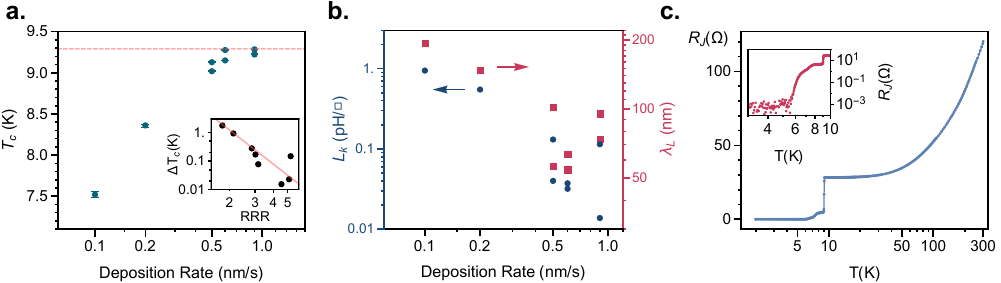}
\caption{
Superconductor material quality. 
(a) Niobium superconducting critical temperature $T_C$ extracted from resistivity measurements as a function of metal deposition rate. At rates above \qty{0.6}{\nm/\s}, $T_C$ approaches bulk value (dashed line). The inset shows deviations from bulk $\Delta T_C = T_C^\text{bulk} - T_C$ are correlated with the residual resistivity ratio, implying high deposition rates result in high-quality films.
(b) Sheet kinetic inductance $L_k$ and observed London penetration depth $\lambda_L$ plotted as a function of deposition rate suggesting that films deposited at higher rates are closer to the clean superconductor limit.
(c) Specific junction resistance $R_J = R/N$ obtained by measuring the resistance $R$ of a chain of $N=12$ junctions as a function of temperature. A sharp drop in resistance is observed above \qty{9}{\K} as the niobium electrodes begin to superconduct. As the temperature decreases, the junction critical currents increase above the excitation current (\qty{10}{\uA}), and below \qty{5}{\K} the measured resistance drops to zero as the excitation is confined to the superconducting branch, indicating proximitization of the aluminum and superconducting contact between the counterelectrode and wiring layers.
\label{fig:figS1}}
\end{figure*}
Josephson junction properties are largely determined by the characteristics of the two superconductors and the insulating oxide barrier that separates them, so the initial formation of the trilayer materials is crucial for the device quality.
As niobium sets the limit of superconducting properties and losses in our junctions and qubits, it is crucial to begin with a high-quality and thus high-purity material.
Maintaining material purity presents a challenge for any thin film deposition technique, made difficult in particular by the incorporation of contaminants into the film during growth.
This contamination can be addressed with two main approaches: first by reducing the flux of contaminants (achieved by reducing the vacuum pressure during the deposition process), but also by reducing the duration of exposure, which can be controlled by the deposition rate.

For electron beam evaporation (the deposition technique used here) vacuum pressures are reduced as low as possible during deposition, however are limited to the \qty{e-8}{\milli\Bar} range by the hardware.
With the contaminant flux fixed by the deposition system vacuum pressure, we explore the effect of deposition rate on Nb purity.
By measuring the resistivity of a film with a known geometry at varying temperatures, we obtain a wealth of information about the film properties.
In Fig. \ref{fig:figS1}a we plot the superconducting transition temperature $T_C$ (proportional to the superconducting gap $\Delta_0$) as a function of metal deposition rate.
We observe that higher rates yield increased transition temperatures, which approach those found in bulk high-purity Nb \cite{williamson1970nbTc}, indicating that the films are increasingly pure.
Indeed, we can also correlate the residual resistivity ratio $RRR=\rho(\qty{300}{\K})/\rho(T_C)$, an indicator of superconductor quality, with deviations of measured critical temperature the bulk value $T_C^\text{bulk}$, supporting the notion that higher deposition rates yield higher-quality films.
Due to the extreme local temperatures required, practical considerations and stability concerns put a limit on feasible deposition rates.
Nonetheless, despite variations induced by vacuum conditions, we find that rates above \qty{0.6}{\nm/\s} are required to deposit a film with high purity.

We can go further to examine the degree of disorder in the superconductor by probing the kinetic sheet inductance $L_K=\hbar R_\Box / \pi \Delta_0$ where $R_\Box=\rho_0/t$ is extracted from the film thickness $t$, and the resistivity just above the superconducting transition.
The sheet inductance also yields the London magnetic penetration depth $\lambda_L^2=t L_K / \mu_0$.
In Fig. \ref{fig:figS1}b we find that both $L_K$ and $\lambda_L$ are also reduced with films deposited at higher rates.
Lower kinetic inductance and shorter London lengths indicate a lower degree of disorder in the superconductor, suggesting that increased deposition rates bring the material further away from the disordered dirty superconductor limit ($\lambda_L \gg \xi$) \cite{lykov1993superconductingstate}.

We verify the superconducting contact quality between the wiring layer and the counter-electrode, as well as the junction tunnel barrier transparency by measuring the voltage accross a chain of 12 junctions in series, through which we send a fixed excitation current of \qty{10}{\uA}.
In Fig. \ref{fig:figS1}c we plot the per-junction specific resistance $R_J$ as a function of temperature, showing the immediately apparent superconducting transition above \qty{9}{\kelvin}.
Immediately below the transition, the superconducting gap is still relatively low, and the junction critical currents fall below the excitation current, so a small resistance is observed.
However as we decrease the temperature, we find that the resistance shrinks by several orders of magnitude (below the noise floor of the instrument).
This indicates that the sum of any remaining resistance channels in a single junction is likely well below the m$\Omega$ range, suggesting a superconducting contact between the Nb wiring layer and the Nb counter-electrode.

\begin{figure*}[ht]
\centering
\includegraphics[width=5.4in]{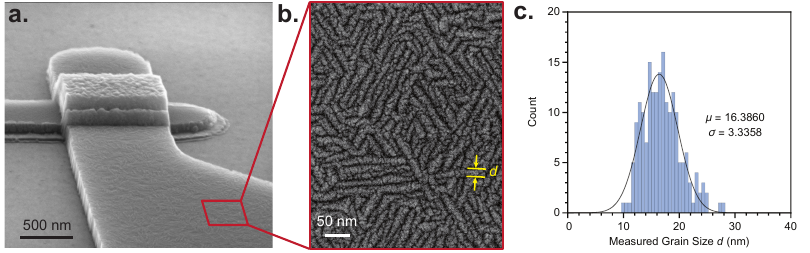}
\caption{
Superconductor grain size. (a) In a tilted scanning electron microscope image of a junction, microscopic grains are observed on the metal surface. In regions of the wiring layer that lie directly on the sapphire substrate, the columnar grain growth is uninterrupted, and the grain pattern is transferred to the top surface of the metal.
(b) A top down high-resolution scanning electron micrograph reveals the hexagonal arrangement of the grains. The grain size can be estimated by measuring the narrow dimension of a grain, marked $d$.
(c) A histogram of repeated measurements of grain width are fitted to a normal distribution which suggests an average grain width of 16.386~nm.
\label{fig:figS1-2}}
\end{figure*}

In a superconductor the residual resistivity ratio is also correlated with grain size in the film \cite{premkumar2021nb,bose2005nbgrain,bose2006nbgrain}.
For a junction wiring layer deposited at 0.9~nm/s we observe a RRR of 4.45, indicating good quality relative to the films we produce (see Fig. \ref{fig:figS1}a).
From the scanning electron microscope image shown in Fig. \ref{fig:figS1-2}a we observe a short-range ordered microscopic grain structure in the regions where the wiring layer is deposited directly on the exposed sapphire substrate.
A high-resolution top-down SEM image shown in Fig. \ref{fig:figS1-2}b reveals a network of thin grains with a visible hexagonal arrangement.
Interestingly since the crystal structure within niobium grains is expected to be cubic \cite{bose2005nbgrain} this suggests the long-range hexagonal order is a reflection of the C-plane sapphire substrate surface.
The individual grains (distinctly larger than the 1-4~nm grains of the Pt and Pd film used to reduce charging in the image) are significantly elongated in one dimension.
To get a sense of the grain size, we measure the short dimension of a grain (as shown in Fig. \ref{fig:figS1-2}b) for a number of grains visible in the image, and summarize the results in Fig. \ref{fig:figS1-2}c.
By fitting to a normal distribution, we find an average grain width of 16.386~nm, with some skew towards longer widths.
Notably we do not see the expected $T_c$ reduction from this grain size \cite{bose2005nbgrain,bose2006nbgrain} since we find the measured $T_c$ for this film is relatively close to the bulk value \cite{finnemore1966nbTc,williamson1970nbTc}.
This suggests that the shortest dimension of the grains does not limit superconductor performance.
We can instead extract an average grain area of $d\times l\simeq1638~\text{nm}^2$ for our film, which corresponds to a effective grain size of $d_\text{eff}=\sqrt{ld}\simeq40.47~$nm, from which we expect properties similar to bulk \cite{bose2005nbgrain,bose2006nbgrain}.
Further investigation using X-ray diffraction or transmission electron microscopy \cite{premkumar2021nb,bose2006nbgrain,verjauw2021nbOxide} could reveal even more details about the microscopic properties of the niobium.

\section{Lossy Plasma Etch Residues}
\label{appendix:c}
\begin{figure*}[ht]
\centering
\includegraphics[width=6.67in]{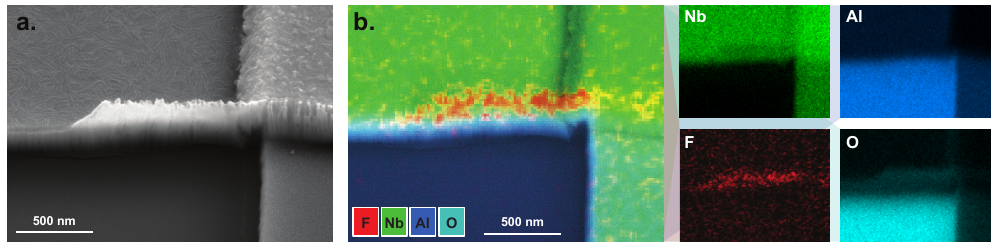}
\caption{
Etch residue chemical analysis. 
(a) Scanning electron micrograph of a plasma etch residue located on the wiring layer near a junction.
(b) Composite Energy Dispersive Spectroscopy (EDS) image overlaid on the image in (a) showing normalized element density regions for F, Nb, Al, and O, with individual element density maps shown in their respective color on the right.
Along with clear Nb and sapphire (\ce{Al2O3}) regions, a high concentration of fluorine relative to the background is found in the residue region, suggesting the residue is composed of fluorinated polymers.
\label{fig:figS2}}
\end{figure*}

By virtue of size, the electric field concentration in a junction is orders of magnitude higher than in the qubit capacitor (or any planar structure such as the resonator capacitor), meaning the participation ratio \cite{wang2015psurf} of the junction side surfaces will also be much higher.
As such, our junction loss is likely still limited by the presence of lossy dielectrics formed on the sides of the junction, which for our design are primarily either spacer material, metal oxides, or residues left by the reactive ion etching process. 
As we cannot use more aggressive spacer \cite{dunsworth2018airbridge} or oxide removal methods \cite{verjauw2021nbOxide} without further risking the integrity of the aluminum junction barrier, we instead study the etch residues and discuss mitigation strategies.

Alongside the desired chemical and mechanical processes that remove niobium, reactive ion etching hosts a variety of simultaneous mechanisms that can grow material:
etched material can either be re-deposited by sputtering, low-energy reaction products can re-adsorb onto exposed surfaces, and components in the plasma can react with exposed material \cite{coburn1979plasmaEtching}.
The products of all of these mechanisms tend to be much more difficult to remove, so end up staying behind after the photoresist is dissolved, particularly on vertical walls not directly exposed to plasma bombardment during the etch.
While the deposited material passivates the walls of the etched region during the etch and can produce high-aspect ratio features, for our junctions its critical to reduce any excess dielectrics, so we explore ways to understand and mitigate these residues in order to reduce loss.

In Fig. \ref{fig:figS2}a we show an example of a dielectric residue located on the side of a junction, which has not been removed throughout the entire fabrication process.
This material must be formed during the third dry etch (Fig. \ref{fig:fig1}f) since it covers and extends off the sides of the Nb wiring and counterelectrode layers.
The residues appear to be present on all vertical surfaces exposed by the etch, visible as striations on the junction sides.
To determine the deposition mechanism for this residue, we probe the chemical composition of the residue using energy dispersive spectroscopy (EDS).
A composite map of normalized element composition is overlaid on the same image of the residue in Fig. \ref{fig:figS2}b, with individual normalized element concentration maps shown to the right.
As expected, we observe high Nb concentrations in the metal regions, and high aluminum and oxygen concentrations in the sapphire region, but more importantly we observe a significant concentration of fluorine in the residue (carbon is also observed in this region as well, but cannot be quantified due to high background carbon levels).
This heavily suggests the residue is some kind of fluorocarbon polymer.

\begin{figure*}[hbt!]
\centering
\includegraphics[width=6.67in]{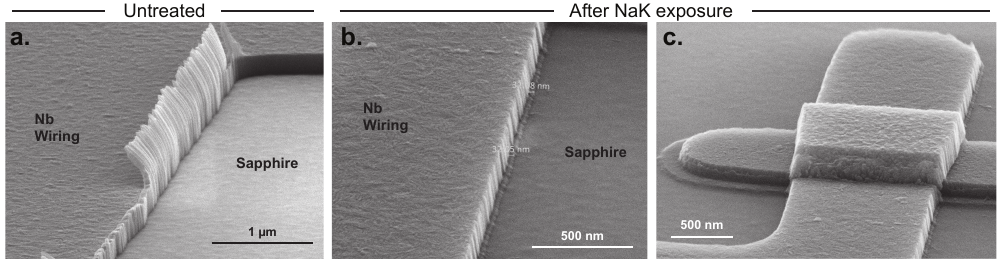}
\caption{
Etch residue NaK reactivity. 
(a) Scanning electron micrograph of a plasma etch residue on the edges of the wiring layer. A closer inspection of the bottom left reveals that the residue extends to cover the sides of the metal, even where the top crust has been mechanically removed.
(b-c) The wiring layer and a junction from the same wafer imaged after a 15 min exposure to sodium-potassium amalgam (NaK) showing nearly complete removal of the etch residue. 
\label{fig:figS2-2}}
\end{figure*}
Fluorocarbons are chemically inert and robust against most standard solvents, acids, or oxygen plasma, and the residues remain largely unaffected by these conditions. However, fluorocarbon polymers are susceptible to defluorination by strong alkali reductants such as sodium-potassium amalgam (NaK) \cite{connelly1996redox,tasker1994ptfe}.
In Fig. \ref{fig:figS2-2}a we show a device with particularly extensive residues covering and extending off the sides of the wiring layer. In an oxygen-free dry nitrogen glovebox, we immerse the sample surface in a sodium-potassium amalgam (NaK) for \qty{15}{\min}, rinse with tetrahydrofuran, move the sample into air, finish rinsing with acetone and isopropanol,  then image the residues.
In Fig. \ref{fig:figS2-2}b-c we observe that the residue material is largely removed: the overhanging features have been removed, as well as the material on metal sides, with the original extent of the residue (about \qty{30}{\nm}) apparent by the indentation left on the sapphire by the residue during the etch.
This corroborates the hypothesis that these residues are composed of fluorocarbons, since the material could be removed upon treatment with NaK, wherein the amalgam cleaves the problematic C-F bonds and allows the remaining residues to become soluble in organic solvents.

While this NaK treatment appears promising on the microscopic scale, in practice the amalgam is difficult to keep clean, and leaves behind significant quantities of dust and salt deposits on the chip surface.
A more practical method to post-clean any residues left behind by the etch might be to instead use a solution with a high reducing potential such as sodium napthalenide \cite{connelly1996redox}, commonly used as a surface treatment for PTFE.
Regardless, the best way to remove the residues is to not form them in the first place, which is achieved by optimizing the etch plasma conditions.
First, we remove obvious residue sources by ensuring the plasma chamber is thoroughly cleaned with oxygen, and no fluorinated vacuum oils are present in the system.
We find that using gas constituents with low hydrogen and carbon content (eg. \ce{SF6} or \ce{CF4}) significantly reduces the residue growth: in particular we find \ce{CHF3} and \ce{C4F8} readily polymerize.
However, we note that using too much \ce{SF6} can lead to the incorporation of sulfur \cite{thomas1987sulfur} into any exposed \ce{SiO2}, which forms an even more inert residue and should be avoided.
The addition of \ce{O2} in the plasma can also help increase the carbon-fluorine ratio of the plasma \cite{mogab1978cf4o2}, but also increases resist etch rate \cite{gokan1983o2etch} and may passivate exposed metal \cite{sasserath1990sloped}, which affects the etch profile.
Using a low density plasma with a long mean free path for the radicals is key to increasing the etch rate and reducing re-deposition, as it increases the effective reactant and product temperature.
Residue formation is also particularly sensitive to substrate temperature.
With the substrate too cold, the reaction product temperature becomes low enough to allow recondensing, leading to increased fluorocarbon deposition.
If the substrate is too hot, reactivity of the photoresist polymers is increased, promoting crosslinking, polymerization, and fluorination: thus good thermal contact between the substrate and the carrier wafer is essential, as the high temperature plasma can otherwise significantly heat the substrate.
Finally, we observe the residue formation accelerates when the insulating substrate is exposed (likely a result of screening charges focusing the plasma towards remaining metal), so we ensure the etch is stopped within \qty{15}{\s} of completion.

\section{Junction Area Dependence, Variation and Stability}
\label{appendix:d}
\begin{figure*}[htb]
\centering
\includegraphics[width=6.67in]{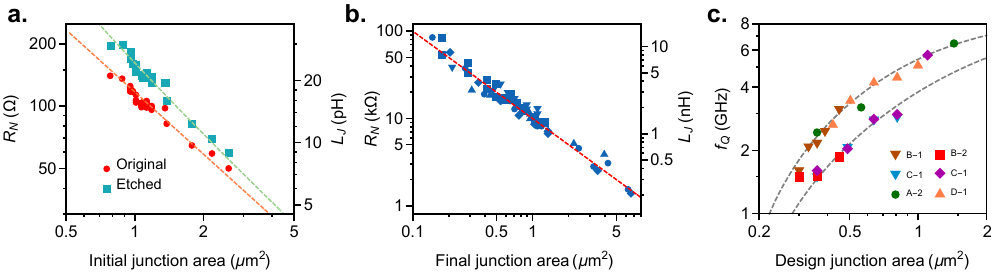}
\caption{
(a) Room temperature junction resistance and junction inductance plotted as a function of junction area (corrected for lithographic reduction).
Original un-treated junction resistances are shown in red, and etched junctions in teal, with fits to an inverse relationship to area (dashed lines) yielding the original critical current density $J_c$ and an etch dimension reduction of approximately \qty{160}{\nm}.
(b) Junction resistances as a function of the final junction area with a inverse fit (dashed line) which gives the critical current density.
For illustrative purposes we have shown PECVD junctions in (a) and HDPCVD junctions in (b).
(c) To estimate reproducibility, spectroscopically measured qubit frequencies are plotted as a function of design junction area, labelled by wafer and cooldown. Expected values for the two different qubit capacitor designs (120 and 160~fF) are shown with dashed lines.
\label{fig:figS3}}
\end{figure*}

Having verified the relationship between the normal state resistance $R_n$, the critical current and the gap energy \cite{ambegaokar1963relation} (see Fig. \ref{fig:fig2}), we can use room temperature resistance measurements to efficiently predict cryogenic junction properties.
In Fig. \ref{fig:figS3}a, we show room temperature junction resistance and junction inductance (calculated from resistance using the $I_c R_N$ product), plotted as a function of junction area (corrected for lithographic reduction).
The original un-treated (see Fig. \ref{fig:fig1}f) junction resistances are in good agreement with the expected inverse dependence on junction area, enabling us to fit the original critical current density.
After etching the spacer (see Fig. \ref{fig:fig1}g) some of the aluminum is removed as well, and the resistance increases since the effective junction dimensions have shrunk.
By fitting the etched junctions, we extract a dimension reduction of approximately \qty{160}{\nm}, which corresponds to about \qty{80}{\nm} of aluminum that gets removed by the etch. 
We note that this sets a practical limit on how small the junction can be before etch effects become more significant than lithographic definition of junction area.

Fitting junction resistances as a function of the final junction area (taking into account the dimension reductions) yields the true critical current density for the final junctions (Fig. \ref{fig:figS3}b). 
We repeat these measurements for wafers with different processing conditions to populate Fig. \ref{fig:fig2}b.
A spread (typically between 5-10\%) is noticeable in our junction resistance for a given junction area.
While higher than typical niobium trilayer junction non-uniformity \cite{Tolpygo2015nbplanarized,bumble2009submicrometer}, our junction variance can primarily be attributed to relatively large geometric deviations due to the limits of our lithographic resolution, which is compounded by fluctuations in the etch dynamics that determine the final junction area.
This implies that our junction parameter spread could likely be reduced with higher resolution lithography methods and a more selective spacer removal technique.
We test the functional limits of our junction reproducibility by measuring deviations of qubit frequencies across different chips from different wafers.
In Fig. \ref{fig:figS3}c, we show spectroscopically measured qubit frequencies (determined by junction inductance) as a function of design qubit junction area for devices with two different qubit capacitor designs.
After determining the qubit capacitance and applying the estimated junction area reductions, we find the measured frequencies are self-consistent within 10 percent or so, even across separate wafers.

\begin{figure*}[htb]
\centering
\includegraphics[width=6.67in]{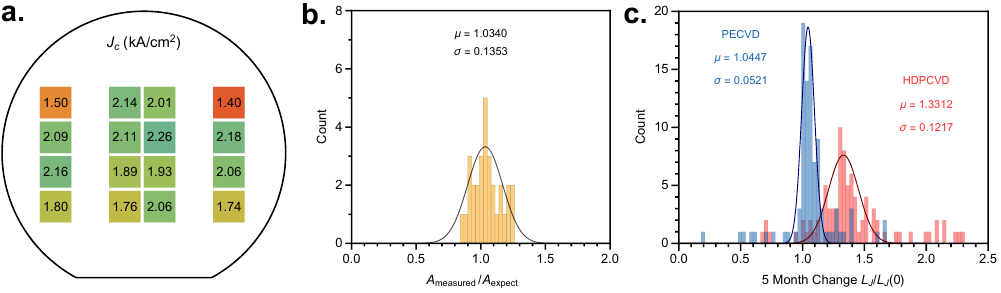}
\caption{
(a) Average junction critical current density on an individual chip measured across several chips across a 2~inch wafer, with deviations from nominal values (2.088~kA/cm$^2$) highlighted with color.
(b) Junction area measured with optical microscopy relative to the expected design area, highlighting the distribution of deviations resulting from lithography.
(c) Long term stability of junctions measured by the relative change in Josephson inductance for 5 month old junctions relative to their original values. Notably the change in high temperature PECVD junctions is much lower than HDPCVD junctions.
\label{fig:figS3-2}}
\end{figure*}

We can investigate the variation of junction properties in more detail by repeating the measurements in Fig. \ref{fig:figS3}b for chips in different physical locations across a 2 inch diameter wafer.
We plot the results by their original position in the wafer and summarize the results in Fig. \ref{fig:figS3-2}a.
We find that the fitted critical current density fluctuates from chip to chip, consistent with the typical 5-10\% junction variation observed in Fig. \ref{fig:figS3}b.
Additionally, we observe a wafer-scale radial dependence in extracted critical current density, with noticeably lower values near the edge of the wafer.
This is likely caused by a combination of dimension deviations from optical lithography and RIE etch rates, both of which have a wafer-scale radial dependence in our process.
To estimate this lithographic dimension variation, we examine the statistics of measured junction area within a single chip relative to the expected area (with critical dimension bias taken into account), summarized in Fig. \ref{fig:figS3-2}b.
Notably the measured areas are distributed with a standard deviation of 13.53\% relative to the expected area:
when accounting for the 5\% accuracy of the area measurement, the remaining spread accounts for a significant amount of the fluctuations observed in junction parameters.
Thus we estimate that the dominant source of junction parameter variation is a result of dimension variation from optical lithography along with further dimension variation from fluctuation in etch dynamics. 
Process uniformity and lithographic dimension conformity are extensively studied topics \cite{Tolpygo2015nbplanarized,bumble2009submicrometer,morohashi1987nbjrev}, so we believe that applying these techniques or moving to higher-resolution lithography (stepper or electron-beam) \cite{kim2021nbn16u,bumble2009submicrometer} could help decrease junction parameter variation.

Josephson junctions are known to change with age \cite{morohashi1987nbjrev,dochev2010aging}, so it is also important to investigate the long-term stability of junction parameters, particularly for our design which leaves the junction barrier exposed from the side.
To this end, we re-measure junctions after 5 months of storage in air.
In Fig. \ref{fig:figS3-2}c we show the relative change in calculated junction inductance for both high-$J_c$ HDPCVD junctions as well as low-$J_c$ PECVD junctions after the storage period.
In both cases we observe an expected increase in $J_c$ from diffusion in the junction barrier with age \cite{morohashi1987nbjrev,dochev2010aging}.
The aged high-temperature PECVD junctions show a mean $J_c$ increase of about 4.5\%, significantly lower than the low-temperature HDPCVD junctions, which exhibit a critical current density increase around 33\% along with a wider distribution.
This suggests that for the high-temperature PECVD junctions, the aging mechanism in the junction barriers is accelerated during the fabrication process.

\section{Junction Annealing Mechanism}
\label{appendix:e}
\begin{figure*}[ht!]
\centering
\includegraphics[width=4.42in]{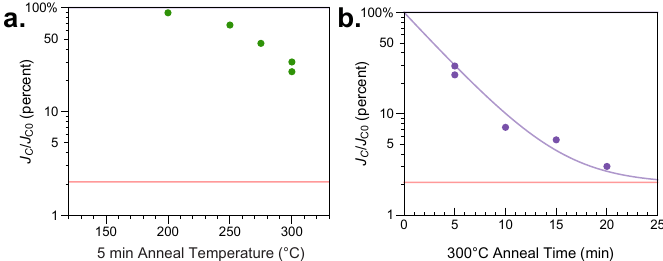}
\caption{
(a) HDPCVD Junction critical current density reduction after annealing for 5 min plotted as a function of anneal temperature showing activation at \qty{250}{\degreeCelsius}.
(b) Critical current density reduction as a function of anneal time at \qty{300}{\degreeCelsius}, which approaches the factor of 50 reduction observed in the main text (red lines).
The purple line represents an exponential fit saturating at the observed reduction factor.
\label{fig:figS6}}
\end{figure*}

The effect of process temperature is readily apparent when comparing the resulting critical current densities of junctions with PECVD spacers (deposited at \qty{300}{\degreeCelsius}) and those with HDPCVD-grown spacers (\qty{90}{\degreeCelsius}).
In Fig. \ref{fig:fig2}b, for the high temperature PECVD junctions, we observed an approximately 97.7\% reduction in $J_c$.
We investigate this effect in more detail by annealing finished low temperature (HDPCVD) junctions with initial $J_{c0}\sim\qty{3}{\kilo\A\per\cm\squared}$ in a dry \ce{Ar} atmosphere, then re-measuring their critical current density.

In Fig \ref{fig:figS6}a, we plot the annealed $J_c$ as a percentage of the untreated $J_{c0}$, and confirm that the annealing effect activates above \qty{200}{\degreeCelsius}, in agreement with \cite{morohashi1987nbjrev}.
In Fig. \ref{fig:figS6}b we show the critical current density of junctions annealed at \qty{300}{\degreeCelsius} for various lengths of time.
After about \qty{20}{\min} (the approximate time wafers spend at \qty{300}{\degreeCelsius} during PECVD) we find that the current density reduction approaches the measured ratio between the PECVD and HDPCVD junctions.
This suggests the high-temperature process dynamically anneals the junction barrier, likely increasing mobility in the oxide barrier which enables diffusion and reduces pinhole density \cite{senapati2010pinhole}.
Qualitatively, this process appears to be exponential in time, so we overlay a saturating exponential fit of the form $J_c/J_c^0 = (1-\alpha)e^{-t/\tau}+\alpha$, where $\alpha$ is the observed reduction factor, and obtain a critical time $\tau \approx \qty{4}{\min}$.
The observed annealing effect is consistent with the critical current densities measured in Ref. \cite{gronberg2017swaps} which do not exceed \qty{190}{\degreeCelsius} during the fabrication process.
With this in mind, our PECVD process could be modified to produce high critical current density junctions by either reducing the deposition temperature below \qty{200}{\degreeCelsius} or to a lesser extent by limiting the time spent at elevated temperatures.
This would allow for improved process stability by providing control over a wide range of critical current densities in a unified process, eliminating the need for switching between PECVD and HDPCVD deposition methods.

\section{Qubit Geometry and Experimental Setup}
\label{appendix:f}
\begin{figure*}[htb]
\centering
\includegraphics[width=5.5in]{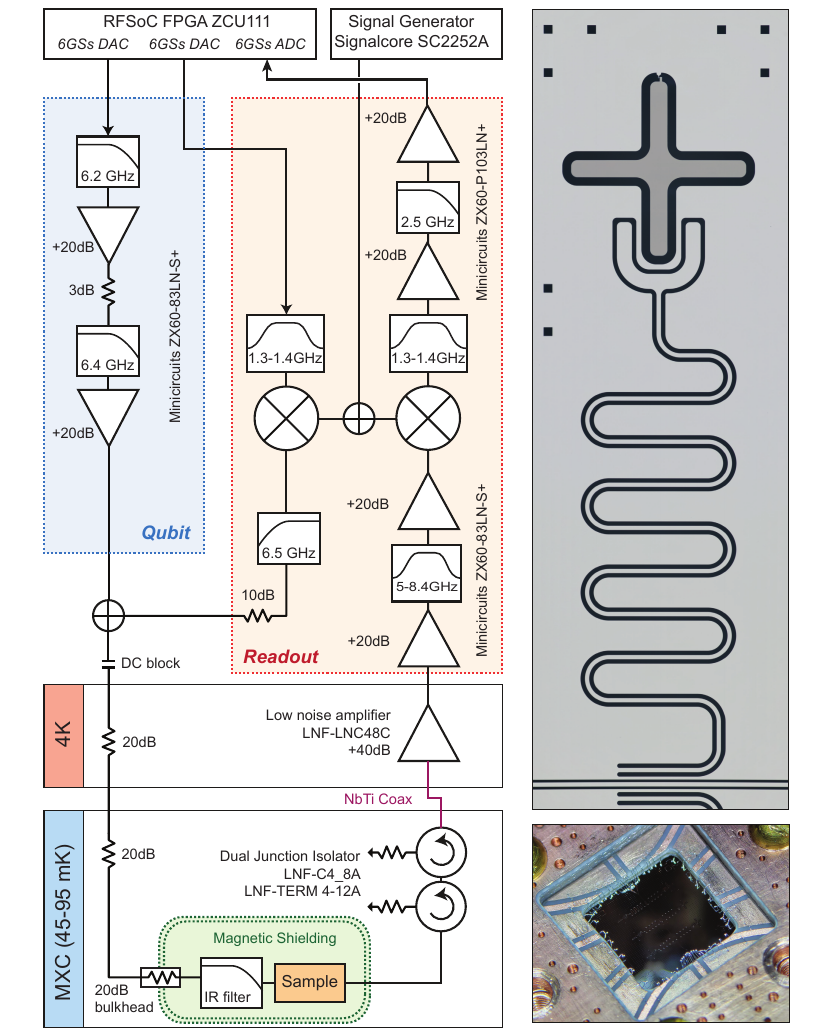}
\caption{
Schematic of the microwave measurement setup used for qubit characterization. Colored tabs show temperature stages inside the dilution refrigerator. 
A composite microscope image (top right) shows a single qubit and its readout resonator, coupled to a waveguide for measurement.
A photograph (bottom right) shows the chip containing 6 qubits mounted in its copper circuit board. 
\label{fig:figS4}}
\end{figure*}

The qubit, readout resonator and other structures are formed in the same steps as the junction.
We base our design on a qubit geometry \cite{barends2013xmon} popular for its reduced radiation profile, a result of the cross-shaped coplanar qubit capacitor whose local electric dipole moments act to cancel each other out far away.
In our case, the cross shape (typically used to implement qubit-qubit coupling or additional charge drives) isn't strictly necessary and a coplanar capacitor composed of any two-dimensional shape would work as well.
We also make an effort to minimize coupling to lossy two-level systems in surface dielectrics by rounding sharp corners where possible in the geometry.
This reduces electric field concentration at specific points in the capacitor, leaving a weaker and more homogenous electric field which should couple less strongly to individual two-level systems.

\begin{figure*}[htb]
\centering
\includegraphics[width=6.67in]{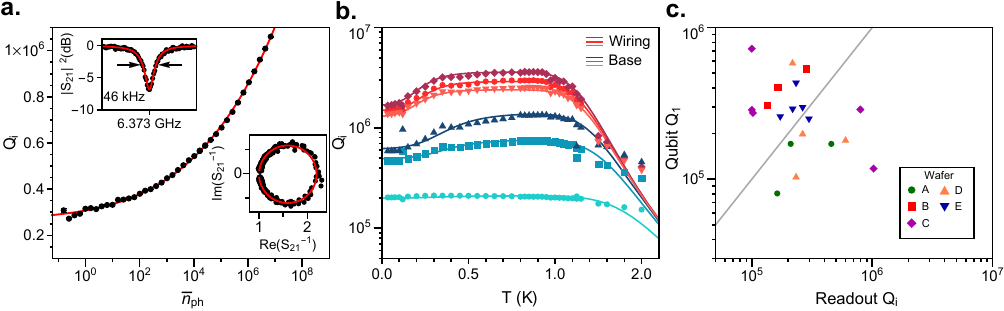}
\caption{
(a) Power dependence of the internal quality factor for a readout resonator ($Q_e=2.6\times10^5$) with no qubit present. The red line is a fit to a model including loss from two-level systems (TLS). The insets show the lineshape and fits at an average photon occupation $\bar{n}_\text{ph}\approx0.96$.
(b) Internal quality factor of resonators without qubits measured as a function of temperature. Solid lines are fits to a model including TLS loss and quasiparticle loss. The three red resonators are formed from the wiring layer, and the blue resonators from the base electrode. Measurements are taken at $\bar{n}_\text{ph}\approx10^4$ so some TLS loss is saturated.
(c) Qubit quality factors $Q_1$ plotted as a function of their readout resonator quality factors $Q_i$ (measured at $n_\text{ph}<1$). A grey line indicates a 1:1 relationship.
\label{fig:figS7}}
\end{figure*}

An example of our qubit geometry is shown in a composite microscope image on the top right of Fig. \ref{fig:figS4}, imaged after Etch 3 (Fig. \ref{fig:fig1}f).
The niobium and un-etched aluminum have visibly different colors, allowing us to distinguish between the wiring layer and the base electrode.
In our geometry, the qubit capacitor is formed with both layers, while the rest of the circuit and the majority of the chip (ground plane, readout resonator and coupling waveguides) is formed with just one layer.
We find that the wiring layer readout resonators exhibit lower loss (See Appendix \ref{appendix:g}), so typically pick the wiring layer for the ground plane.
However having measured devices with both configurations (majority wiring layer and majority base electrode), we don't find extreme differences in qubit properties, where the fields participate in both layers regardless of orientation.
As an example, compare base-electrode ground plane wafer D with wiring ground plane wafer A in Fig. \ref{fig:figS7}b, whose quality factors are similar.

The qubits are capacitvely coupled to a meandered quarter wave coplanar waveguide resonator, which is in turn inductively coupled to a transmission line for readout.
For simplicity, we couple directly to the readout resonator without additional purcell filtering.
Chips containing up to 6 qubits and resonators are mounted in a copper circuit board shown in the bottom right of Fig \ref{fig:figS4}, which is in turn bolted to a copper post thermalizing the assembly to the base temperature of an Oxford Triton 200 dilution refrigerator with minimum mixing chamber temperatures between \qtyrange[range-units = single,range-phrase = --]{45}{95}{\milli\K}.
The mounted assembly is encased in two layers of Mu-metal magnetic shielding to reduce decoherence from stray magnetic fields, the qubits are isolated from microwave noise through an Eccosorb CR-110 high-frequency absorbing filter as 60 dB of cryogenic attenuation which keep the input noise close to the mixing chamber temperature.
Transmitted microwave signals pass through two wideband circulators (isolating the qubits from microwave noise from the output side) into a low-loss superconducting NbTi coaxial cable, then are amplified by a low noise cryogenic amplifier followed by additional room temperature amplification.

Resonators and qubit transitions are characterized with single and two-tone spectroscopy using a Agilent E5071C network analyzer.
For pulsed qubit measurements, we use a Quantum Instrument Control Kit \cite{stefanazzi2022qick} based on the Xilinx RFSoC ZCU111 FPGA.
Qubit pulses are directly synthesized by the FPGA, while measurement pulses are generated with a heterodyne conversion setup, as shown in Fig. \ref{fig:figS4}.
With the spectral layout of each device determined, we select filter networks to minimize unwanted images and harmonics from the FPGA for both the qubit and readout pulses, with a broadband example configuration shown in Fig. \ref{fig:figS4}.
The FPGA and carrier signal generator are clocked to a 10 MHz rubidium source for frequency stability.

\section{Material Loss Probed by Resonator Quality Factor}
\label{appendix:g}

To compare qubit loss contributions from material sources with contributions from the junction itself, we measure quality factors for readout resonators subject to the same fabrication conditions, but with no qubits attached.
A typical normalized transmission spectrum of a resonator taken at a low average photon number ($\bar{n}_{\text{ph}}\approx0.96$) is shown in the inset of Fig. \ref{fig:figS7}a. 
On resonance, we observe a dip in magnitude, which at low powers is described well by \cite{khalil2012analysis}:
\begin{equation}
    S_{21}=1-\frac{Q}{Q_e^*}\frac{e^{i\phi}}{1 + 2i Q \frac{\omega-\omega_0}{\omega_0}}
    \label{eq:lowpowerS}
\end{equation}where $Q^{-1}=Q_i^{-1}+\text{Re}[Q_e^{-1}]$ and the coupling quality factor $Q_e= Q_e^* e^{-i\phi}$ has undergone a complex rotation $\phi$ due to minor impedance mismatches.
We plot fitted internal quality factors in Fig. \ref{fig:figS7}a, finding that $Q_i$ increases with power.
This behavior is entirely captured by a power dependent saturation mechanism \cite{wang2009improvingtls}, suggesting the dominant loss mechanism in the resonators arises from coupling to two-level systems.

We further investigate limits on the resonator loss by using increased temperatures to further saturate the two-level systems.
In Fig. \ref{fig:figS7}b we plot $Q_i$ measured at $\bar{n}_{\text{ph}}\approx10^4$ as a function of temperature (grouped by fabrication layer),
with solid lines corresponding to a model of the form
\begin{equation}
    Q_i(T)^{-1}=Q_{\text{other}}^{-1} + Q_{\text{TLS}}(T)^{-1} + Q_{\sigma}(T)^{-1}
\end{equation}
where $Q_{\text{TLS}}$ is the saturating loss mechanism from two-level systems \cite{wang2009improvingtls}, $Q_{\text{other}}$ is a temperature independent upper bound arising from other sources of loss, and the conduction loss $Q_\sigma$ is given by \cite{mattis1958bardeen}:
\begin{equation}
    Q_\sigma(T)=\frac{1}{\alpha}\frac{\sigma_2(T,T_c)}{\sigma_1(T,T_c)}
\end{equation}
where $\sigma_1$ and $\sigma_2$ are the real and imaginary parts respectively of the complex surface impedance, calculated by numerically integrating the Mattis-Bardeen equations for $\sigma_1/\sigma_n$ and $\sigma_2/\sigma_n$ \cite{mattis1958bardeen}. $T_c$ is constrained to the values measured in Appendix \ref{appendix:b}, and $\alpha$ is used as a fit parameter.

Comparing resonators formed during different steps in the fabrication process, we observe that resonators made from the wiring layer exhibit consistently higher quality factors, while resonators from the base layer are lossier and much more variable.
Since the sides of the base layer have been exposed to more fabrication steps than the wiring layer, the surface niobium of this layer has a much longer chance to oxidize, and has the additional potential to host lossy dielectrics from un-removed spacer material.
Thus, while we have improved losses in the wiring layer to about $Q_\text{TLS} \sim 0.9\times 10^6$ by reducing fluorocarbon formation, our devices are still loss-limited to approximately $2\times 10^5$ by two-level systems in the surfaces of the base electrode.

To investigate the relationship between qubit and readout resonator decoherence, we also measure quality factors of the readout resonator for each qubit.
At single-photon powers, the readout resonator is maximally susceptible to material-based loss from two-level systems in its surface, but due to the hybridization of its electric field with the qubit mode will also interact with the materials in the qubit.
In Fig. \ref{fig:figS7}c we compare qubit quality factors $Q_1$ with the single-photon readout quality factor $Q_i$ for each of the devices from Fig. \ref{fig:fig3}.
On average, we observe that the two quality factors are close to a one to one relationship (as indicated by the grey line), with device variations within a factor of 3 or so. 
While a direct correlation between the two cannot be extracted from this data, this is to be expected for loss dominated by inhomogeneous material defect distributions between the resonator and qubit. 
Nevertheless, the similarity of the two quality factors leads us to conclude the qubit and resonator are likely limited by similar decoherence mechanisms.

\section{Detailed Model of Junction Losses}

\begin{figure*}[htb]
\centering
\includegraphics[width=6.67in]{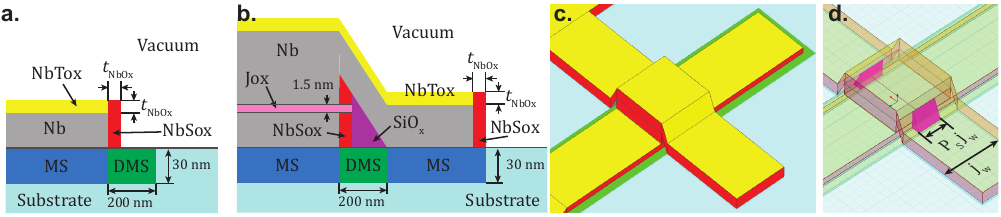}
\caption{
Junction loss regions (a) Cartoon showing regions defined for a resonator made with the first layer, with dimensions exaggerated.
Niobium oxide (metal-air interface) is separated into top oxide (Tox) and side oxide (Sox) regions.
For a wiring layer resonator, the dirty substrate region (DMS) is merged with the substrate layer.
(b) Cartoon showing regions for a junction, which adds the junction barrier region (Jox) and the spacer region (SiOx).
(c) Three dimensional rendering of the junction with realistic dimensions. Simulated regions are colored in the same way as in parts (a-b).
(d) Transparent rendering of the junction visualizing the spacer remaining percentage $P_S$ relative to the junction width $j_w$.
\label{fig:figS8-1}}
\end{figure*}

In the main text along with Appendix \ref{appendix:g} we established that our junction quality factor ($Q_J\approx10^5$) is similar to the single-photon quality factors of bare resonators, for which we measured an average of $2.6\times10^5$ for base layer and $1.04\times10^6$ for the wiring layer.
The fact that these loss rates are comparable suggest that some part of the qubit decoherence arises from same material losses probed by the resonators.
To investigate the origins of these loss channels in more detail and elucidate important pathways for further improvement, we use finite element method simulations (Ansys HFSS) to examine the energy participation ratios \cite{wang2015psurf} of different regions and interfaces in the junction.

\begin{figure*}[hbt]
\centering
\includegraphics[width=6.67in]{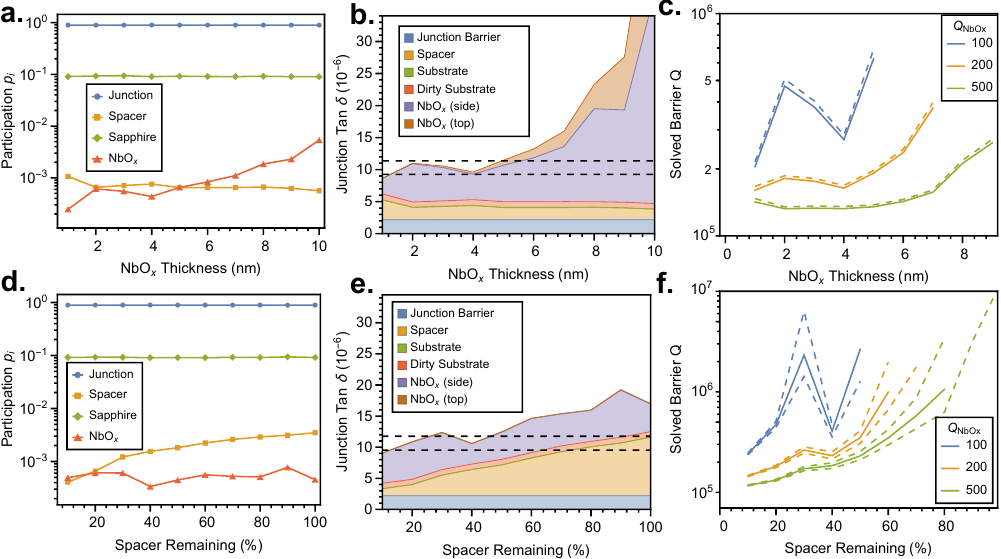}
\caption{
Junction losses by region.
(a) Participation ratios of the primary lossy materials in the junction, plotted as a function of niobium oxide thickness $t_{\text{NbO}_x}$. As expected the niobium oxide participation ratio increases as the layer gets thicker.
(b) Junction loss tangent expressed as visual sum of losses from various materials in the junction with assumed loss tangents, plotted as a function of niobium oxide thickness.
For thicker oxide layers (eg. those used in anodization processes) niobium oxide loss dominates the junction loss.
The junction loss calculated from Fig. \ref{fig:fig4}c is shown in black dashed lines. 
(c) We can also solve for the barrier quality factor based on the junction quality factor and the calculated participation ratios for varying material quality factors.
Solid and dashed lines correspond to a \ce{SiO2} loss tangent of $\tan \delta=2.7\times10^{-3}$ and $2.9\times10^{-3}$ respectively.
In (d-f) we repeat parts (a-c) but measure the effect of partially un-removed spacer material expressed as a fraction $P_S$ of the junction width. 
We find that residual spacer material contributes a significant amount of loss.
For both sets of simulations, the unswept variable is set to nominal values of $t_{\ce{NbOx}}=2~$nm and $P_S=0.2$.
\label{fig:figS8-2}}
\end{figure*}
Figure \ref{fig:figS8-1} illustrates the material regions studied.
Similar to other surface participation studies \cite{wang2009improvingtls,wang2015psurf,Dial2016particip} we consider the metal-substrate interface regions, which we further subdivide into the metal-substrate interface (MS) and the dirty metal-substrate region (DMS) which may contain some remaining spacer material.
As the etched sapphire surface and bulk loss are both expected to be minimal \cite{Read2023dipper} we combine the substrate-air (SA) interface with the substrate region for participation calculations.
Based on the electric field density we choose the thickness of the surface regions of the substrate to be 30~nm: adjusting this thickness will simply rescale the effective participation and loss of the metal substrate regions.
The bulk of the loss is expected to lie in the amorphous oxide dielectric regions of the junction.
These can be separated into the aluminum oxide comprising the junction barrier (Jox) which we expect to be 1--2~nm \cite{morohashi1987nbjrev}, and the niobium oxide, which we further separate into a top oxide layer (NbTox) and side oxide (NbSox): the latter of these should be substantially lossier since it may contain fluorocarbons after exposure to the fluorine plasma (see Appendix \ref{appendix:c}).
Finally we also consider the possibility of incomplete spacer removal and also include a portion of \ce{SiO2} to model the spacer, as shown in Fig. \ref{fig:figS8-1}d.
With our imaging methods, we are unable to determine the amount or layout of the residual spacer material, so for simplicity we approximate the region as a uniform percentage of the original spacer volume $P_S\leq1$.

Integrating the simulated electric fields in the junction geometry determines the participation ratios in each dielectric region \cite{wang2015psurf}.
The surface niobium oxide thickness $t_{\text{NbO}_x}$ typically ranges between 1--5~nm \cite{verjauw2021nbOxide,Altoe2022NbOx} but can vary to a greater depending on process conditions, so in Fig. \ref{fig:figS8-2}a we study the participation ratios as a function of $t_{\text{NbO}_x}$.
From this we conclude that most of the energy is stored in the junction barrier, followed by the sapphire regions, with the niobium oxide and spacer regions contributing less than one percent.
As expected, the niobium oxide participation increases with $t_{\text{NbO}_x}$, however importantly the energy participation is dominated by the side oxide ($p_\text{Sox}\gg p_\text{Tox}$), especially for thinner values of $t_{\text{NbO}_x}$.
In the same manner, we can also simulate the participation ratios of a section of coplanar waveguide (the cross section of which will be the same as that shown in Fig. \ref{fig:figS8-2}a).
Comparing resonators fabricated from the first and wiring layer effectively amounts to the presence of the dirty substrate region (DMS) in our model.
For the resonator geometry, we find this region has an average participation $p_\text{DMS}=0.366\%$, largely independent of oxide thickness (for which $p_{\text{NbO}_x}\sim0.005\%$, similar to Ref. \cite{verjauw2021nbOxide}).
Based on the single-photon resonator quality factors from both layers in Appendix \ref{appendix:g}, we solve for the material quality factors as a function of the niobium oxide quality factor, which is typically $Q_{\text{NbO}_x}=1/\tan\delta_{\text{NbO}_x} \simeq 100$ \cite{verjauw2021nbOxide,Altoe2022NbOx}.
From this we conclude that $Q_\text{DMS}\simeq1.4\times10^3$ and $Q_\text{Sapphire}\simeq1.8\times10^6$, which is reasonably consistent for averaged bulk and surface measurements of sapphire loss \cite{Read2023dipper} and closer to the loss values found in silicon oxide ($\tan\delta_{\ce{SiO2}} \simeq 2.8\times10^{-3}$\cite{oconnell2008dielectrics}) in the DMS region.

Combining the expected material losses with the calculated participation ratios, we can express the junction quality factor as a sum of loss contributions from each region to identify dominant sources of decoherence.
\begin{equation}
    \tan\delta_J = \frac{1}{Q_J} = \sum_x \frac{p_x}{Q_x}=\sum_x p_x \tan \delta_x
\end{equation}
We summarize the contributions for each material as a function of $t_{\text{NbO}_x}=2~$nm in Fig. \ref{fig:figS8-2}b along with the average junction quality factors measured in the main text.
Despite the high barrier participation, we find the dominant loss contribution is from the niobium oxide: specifically that on the sides of the metal (NbSOx) which is also more likely to be impacted by the plasma etch chemistry. 
For simplicity, we have determined the junction barrier quality factor from the average junction quality factor $Q_J$ by assuming that $t_{\text{NbO}_x}=2~$nm \cite{verjauw2021nbOxide} and conservatively estimating $P_S=0.2$: this yields a junction barrier oxide quality factor of $Q_{\text{Jox}} \simeq 4.7\times10^5$.
We can repeat this calculation with varying conditions to estimate the effective barrier quality, as shown in Fig. \ref{fig:figS8-2}c which suggests that the barrier Q may exceed our estimate if the oxide thickness is thicker than 2~nm, or may be lower if the niobium oxide quality factor is in fact higher than expected.
As the predicted junction loss cannot exceed the measured value, assuming standard oxide loss $Q_{\text{NbO}_x}=100$ \cite{verjauw2021nbOxide,Altoe2022NbOx} implies $t_{\text{NbO}_x}<5~$nm, which helps validate the previous assumptions.

We can also perform a similar set of calculations for the remaining spacer amount $P_S$, summarized in Fig. \ref{fig:figS8-2}d-f.
As expected we observe the silicon participation ratio $p_{\ce{SiO2}}$ increases with larger spacer volume.
When more than half of the spacer remains, we estimate that the silicon oxide comprises the dominant source of loss in the junction.
Similar to the niobium oxide thickness, the spacer percentage also affects the estimated junction barrier $Q$ as shown in Fig. \ref{fig:figS8-2}f.
This also suggests an upper bound for the residual spacer percentage $P_S\lesssim0.5$, indicating the final wet etch is at least somewhat successful in removing spacer material under the wiring layer.

Thus we have identified several key areas where junction loss could be further improved.
As discussed in the main text, reducing the amount of lossy dielectrics (particularly the spacer material and niobium oxide) is key to increasing junction loss, as highlighted in Fig. \ref{fig:figS8-2}b,e.
While we have taken steps to reduce the volume of both niobium oxides and spacer material, further improvements on both these fronts could help improve junction quality.
Further reduction of junction loss may require addressing losses in the dirty substrate region with improved cleaning methods.
However from Fig. \ref{fig:figS8-2}a,c we conclude the junction is most sensitive to the quality of the barrier dielectric.
In this regard, atomically uniform barriers such as AlN deposited with molecular beam epitaxy in NbN junctions \cite{kim2021nbn16u,yu2002nbn4u} may provide even better performance.

\bibliography{thebibliography}
\end{document}